\documentclass[%
 reprint,
 superscriptaddress,
 amsmath,amssymb,
 aps,
 prx,
]{revtex4-2}

\usepackage{graphicx}
\usepackage{dcolumn}
\usepackage{bm}

\usepackage{upgreek}

\usepackage{makecell} 

\def\DDDD{$D_{3d}$ }

\newcommand{\nv}{NV\textsuperscript{--}}
\newcommand{\siv}{SiV\textsuperscript{--}}
\newcommand{\hsiv}{\textsuperscript{29}SiV\textsuperscript{--}}
\newcommand{\gev}{GeV\textsuperscript{--}}
\newcommand{\snv}{SnV\textsuperscript{--}}
\newcommand{\hsnv}{\textsuperscript{117}SnV\textsuperscript{--}}
\newcommand{\cspin}{\textsuperscript{13}C}

\def\ket#1{\left|{#1}\right\rangle}

\def\melement#1#2#3{\left\langle{#1}\left|{#2}\right|{#3}\right\rangle}

\def\_#1{_\mathrm{#1}}
\def\^#1{^\mathrm{#1}}

\def\ud{\uparrow/\downarrow}
\def\UD{\Uparrow/\Downarrow}

\def\purple{$0\_B0\_M \leftrightarrow 0\_B1\_M$}
\def\blue{$0\_B0\_M \leftrightarrow 1\_B0\_M$}
\def\green{$0\_B1\_M \leftrightarrow 1\_B1\_M$}

\def\lambdaeff{\lambda_\mathrm{eff}}

\def\ttwostar{T_2^*}

\begin{document}

\preprint{APS/123-QED}

\title{High-Fidelity Control of a Strongly Coupled Electro-Nuclear Spin-Photon Interface}

\author{Isaac B. W. Harris} 
\thanks{These authors contributed equally to this work.}
\affiliation{Department of Electrical Engineering and Computer Science, Massachusetts Institute of Technology, Cambridge, Massachusetts 02139, USA}

\author{Ian Christen} 
\thanks{These authors contributed equally to this work.}
\affiliation{Department of Electrical Engineering and Computer Science, Massachusetts Institute of Technology, Cambridge, Massachusetts 02139, USA}

\author{Sofia M. Patom\"aki} 
\affiliation{Department of Electrical Engineering and Computer Science, Massachusetts Institute of Technology, Cambridge, Massachusetts 02139, USA}

\author{Hamza Raniwala} 
\affiliation{Department of Electrical Engineering and Computer Science, Massachusetts Institute of Technology, Cambridge, Massachusetts 02139, USA}

\author{Maxim Sirotin} 
\affiliation{Department of Electrical Engineering and Computer Science, Massachusetts Institute of Technology, Cambridge, Massachusetts 02139, USA}

\author{Marco Colangelo} 
\affiliation{Department of Electrical Engineering and Computer Science, Massachusetts Institute of Technology, Cambridge, Massachusetts 02139, USA}

\author{Kevin C. Chen} 
\affiliation{Department of Electrical Engineering and Computer Science, Massachusetts Institute of Technology, Cambridge, Massachusetts 02139, USA}

\author{Carlos Errando-Herranz} 
\affiliation{Department of Electrical Engineering and Computer Science, Massachusetts Institute of Technology, Cambridge, Massachusetts 02139, USA}
\affiliation{Institute of Physics, University of M\"unster, 48149, M\"unster, Germany}


\author{David J. Starling}
\affiliation{Lincoln Laboratory, Massachusetts Institute of Technology, Lexington, Massachusetts 02421, USA}

\author{Ryan Murphy}
\affiliation{Lincoln Laboratory, Massachusetts Institute of Technology, Lexington, Massachusetts 02421, USA}

\author{Katia Shtyrkova}
\affiliation{Lincoln Laboratory, Massachusetts Institute of Technology, Lexington, Massachusetts 02421, USA}

\author{Owen Medeiros} 
\affiliation{Department of Electrical Engineering and Computer Science, Massachusetts Institute of Technology, Cambridge, Massachusetts 02139, USA}

\author{Matthew E. Trusheim}
\affiliation{DEVCOM, Army Research Laboratory, Adelphi, MD, 20783, USA}
\affiliation{Department of Electrical Engineering and Computer Science, Massachusetts Institute of Technology, Cambridge, Massachusetts 02139, USA}

\author{Karl Berggren}
\affiliation{Department of Electrical Engineering and Computer Science, Massachusetts Institute of Technology, Cambridge, Massachusetts 02139, USA}

\author{P. Benjamin Dixon}
\affiliation{Lincoln Laboratory, Massachusetts Institute of Technology, Lexington, Massachusetts 02421, USA}

\author{Dirk Englund} 
\affiliation{Department of Electrical Engineering and Computer Science, Massachusetts Institute of Technology, Cambridge, Massachusetts 02139, USA}

\date{\today}

\begin{abstract}
Long distance quantum networking requires combining efficient spin-photon interfaces with long-lived local memories.
Group-IV color centers in diamond (\siv{}, \gev{}, and \snv{}) are promising candidates for this application, containing an electronic spin-photon interface and dopant nuclear spin memory.
Recent work has demonstrated state-of-the-art performance in spin-photon coupling and spin-spin entanglement.
However, coupling between the electron and nuclear spins introduces a phase kickback during optical excitation that limits the utility of the nuclear memory.
Here, we propose using the large hyperfine coupling of \hsnv{} to operate the device at zero magnetic field in a regime where the memory is insensitive to optical excitation.
We further demonstrate ground state spin control of a \hsnv{} color center integrated in a photonic integrated circuit, showing 97.8\% gate fidelity and 2.5~ms coherence time for the memory spin level.
This shows the viability of the zero-field protocol for high fidelity operation, and lays the groundwork for building quantum network nodes with \hsnv{} devices.

\end{abstract}

\maketitle

\section{Introduction}\label{sec:intro} 

Diamond color centers in photonic structures are a compelling platform for quantum networking~\cite{Bernien2013, Kalb2017a, Pompili2021, Stas2022}, offering coupling between stationary electron spin qubits and optical photons to perform remote entanglement~\cite{Barrett2005}.
While initial research has focused on the \nv{} center, producing state-of-the-art quantum networking demonstrations~\cite{Bernien2013,Hensen2015,Humphreys2018}, current research has moved towards group-IV vacancy color centers (\siv{}, \gev{}, \snv{}) due to their insensitivity to electric field noise~\cite{DeSantis2021} and superior optical coherence when integrated in nanostructures~\cite{Sukachev2017, Wan2020, ArjonaMartinez2022}.
These advances have allowed demonstrations of heralded entanglement between nanostructure-integrated color centers~\cite{Knaut2024}.

Despite progress in generating heralded entanglement with group-IV color centers, network losses still limit the rates at which high fidelity quantum entanglement can be shared using these protocols.
The limits of heralded entanglement can be overcome by adding a local memory qubit to the color center which can store quantum information during repeated entanglement attempts.
After entanglement generation, the stored information can then be used to improve the fidelity of the shared state through entanglement distillation~\cite{Bennett1996, Campbell2008, Kalb2017a} and quantum repeater protocols~\cite{Childress2005, Pompili2021}, or to go beyond single entanglement links to generate large entangled cluster states~\cite{Benjamin2006,Nickerson2014,Choi2019}.
In these schemes, the electron spin serves as an entanglement ``broker'', repeatedly attempting heralded remote entanglement through the lossy network, while high fidelity information is stored in the local memory.
Here we introduce and demonstrate the tin-vacancy (\snv{}) color center strongly coupled to a $^{117}$Sn nuclear memory as brokered entanglement node.
We first motivate the need for this system, and then show that satisfies the requirements for brokered entanglement, to be able to:
\begin{enumerate}
\item read-out (Sec.~\ref{sec:optical}) and attempt heralded entanglement on the broker qubit without disturbing the information stored in the memory.
\item coherently and independently manipulate the broker and memory qubits (Sec.~\ref{sec:fidelity}), 
\item perform local entangling operations between the broker and memory qubits (Sec.~\ref{sec:spin}).
\end{enumerate}

\subsection{Optical Phase Kickback}

\begin{figure*}[ht]
    \centering
    \includegraphics[scale=0.75]{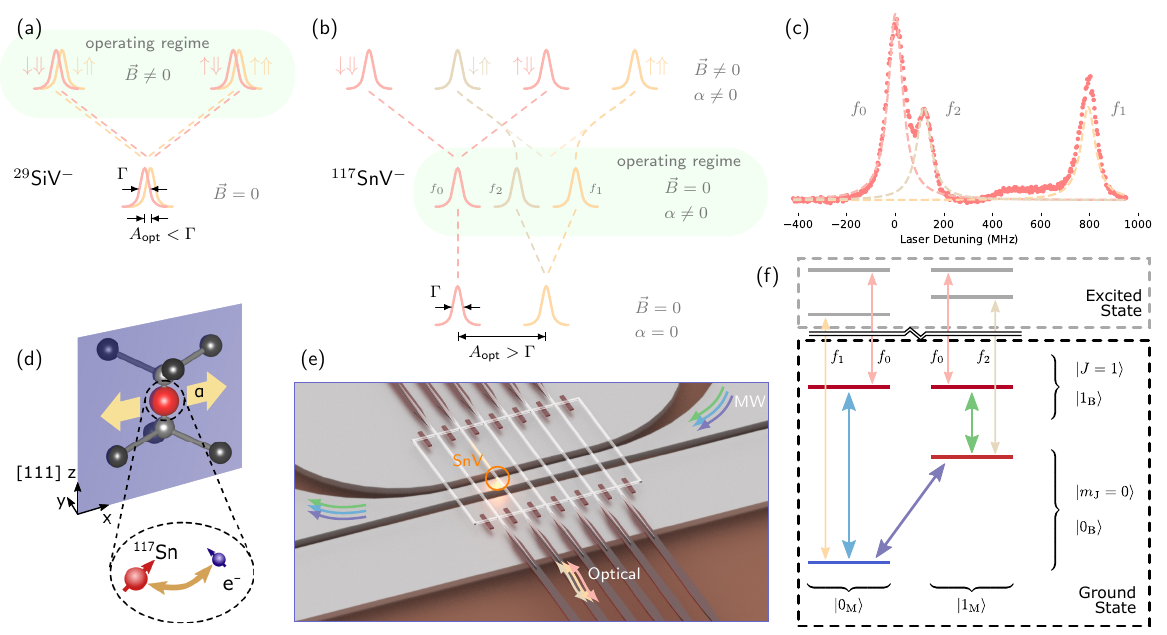}
    \caption{
    (a) While $^{29}$SiV$^-$ requires a magnetic field to optically resolve transitions,
    (b) the optical hyperfine splitting $A_\text{opt}$ of \hsnv{} is much larger than the optical linewidth $\Gamma$, which permits operation in the zero-field regime. In the presence of strain $\alpha$, this produces a characteristic 3-peak structure, observable in 
    (c) the photoluminescence excitation spectrum.
    Peaks $f_0$, $f_1$, and $f_2$ correspond to the transitions labeled in panel f.
    (d) The split-vacancy structure the \hsnv{}.
    The group-IV \textsuperscript{117}Sn atom is shown in red, the nearest-neighbor carbon atoms in black, and the lattice vacancies in gray.
    (e) These \hsnv{} artificial atoms are integrated onto a photonic circuit by stamping diamond waveguides to couple with the silicon nitride waveguides.
    An aluminum microwave line runs underneath the diamond to manipulate the ground-state manifold, as shown in 
    (f) the diagram of the level structure, highlighting the delineation between the broker and memory qubit, as well as the relevant allowed optical  and microwave transitions.
    }
    \label{fig:intro}
\end{figure*}

While using a nuclear spin as a memory qubit is a well-established approach, independent readout and entanglement of the broker while maintaining coherent information in the memory remains challenging.
Since the memory qubit's Larmor frequency can change under optical excitation due to differences in the hyperfine interaction between the ground and excited state, optical excitation for readout or entanglement attempts can lead to decoherence.
As shown in Appendix~\ref{sec:optical_insensitivity}, if the memory is prepared in an initial superposition state $\ket{\psi\_{M}} = \left(\ket{0\_{M}}+\ket{1\_{M}}\right)/\sqrt{2}$, then the fidelity after $n$ optical excitations with mean lifetime $\tau$ (for readout or heralded entanglement attempts) will be
\begin{equation}\label{eq:excitation_fidelity}
    F = \frac{1}{2}\left( 1 + (1 + \Delta\omega\_{M}^2\tau^2)^{-n/2} \right),
\end{equation}
where $\Delta\omega\_{M}$ is the change in local memory frequency under optical excitation.
The fidelity of the memory state decays exponentially in the number of optical excitations so long as $\Delta\omega\_{M}\tau\neq 0$.
As the number of excitation attempts is related to the collection efficiency $\eta$ by $\eta\approx n^{-1}$, and $\eta \ll 1$ in realistic networks, it is imperative to make $\Delta\omega\_{M}\tau$ as small as possible to minimize this source of decoherence.

Previous brokered entanglement demonstrations with the \nv{} system have addressed this problem by optically exciting the $m_S=0$ state of the spin-1 electronic broker qubit, with the intrinsic \textsuperscript{14}N nuclear spin serving as the memory qubit~\cite{Bernien2013}.
The $m_S=0$ state has the attractive property that the hyperfine interaction in this electronic state is identically zero in both the ground and excited electronic states, ensuring $\Delta\omega\_M=0$, and therefore that readout and entanglement operations on the broker do not disturb the memory.
In contrast, the spin-$1/2$ negatively-charged group-IV color centers do no have an equivalent to the $m_S=0$ electronic state with zero hyperfine interaction.
While quantum operations in strongly coupled intrinsic \textsuperscript{29}Si memories have been demonstrated~\cite{Stas2022, Knaut2024}, they have only been used to store states and enhance fidelity of heralded entanglement, not for entanglement distillation or repeater protocols.
Limiting brokered entanglement memories to weakly coupled spins---such as proximal \cspin{}---has also been suggested, since weak coupling limits the magnitude of $\Delta\omega\_M$~\cite{Nguyen2019, Nguyen2019a}.
However, excitation-induced decoherence may still remain large for \cspin{}~\cite{Beukers2024a}, \cspin{} nuclear spin operations are much slower than electron spin operations, and the inclusion of \cspin{} with the \snv{} is not deterministic.

\subsection{\hsnv{} Zero-Field Bias Protocol}\label{sec:theory}

To resolve this problem, we introduce a quantum network protocol based on the \hsnv{} color center in diamond, where the built-in $^{117}$Sn nucleus acts as an intrinsic local memory.
Whereas the weak hyperfine coupling to \textsuperscript{29}Si or \textsuperscript{13}C memories means that a magnetic field must be applied for optical accessibility (Fig.~\ref{fig:intro}a),
in \hsnv{}, the frequencies of the hyperfine optical transitions are separated by an amount ($A\_{opt}$) larger than the optical linewidth, $\Gamma$, even at zero magnetic field (Fig.~\ref{fig:intro}b, Appendix~\ref{sec:hyperfine_hamiltonian}).
We show theoretically that this property allows the \hsnv{} memory operating at zero field to be insensitive to optical excitation.
We further experimentally demonstrate that we can fully control the broker and memory qubit with high fidelity at the moderate temperature of 1.3~K under this regime.
This zero-field protocol operating above 1~K represents a significant step towards the deployment of color-center-based quantum network nodes, free of the constraints of bulky high-field magnets and dilution refrigerator temperatures.

Our protocol depends on the remarkably-large hyperfine coupling of \hsnv{}.
Group-IV color centers exhibit a trend of stronger hyperfine coupling with their intrinsic dopant as the dopant becomes heavier~\cite{Harris2023}.
In the case of \hsnv{}, the difference between the large hyperfine coupling in the ground and excited states is larger than the lifetime-limited optical linewidth, allowing the hyperfine levels to be spectrally resolved without applying any additional magnetic field bias, a feature unique among leading color centers.
These hyperfine levels may also be used to store information~\cite{Parker2023}, a feature which we will extend here to demonstrate a viable group-IV quantum networking node for brokered entanglement meeting all the criteria outlined in Section~\ref{sec:intro}.

As derived in Appendix~\ref{sec:hyperfine_hamiltonian}, the ground state manifold of the \hsnv{} has energy levels
\begin{equation}\label{eq:levels}
    \begin{split}
        E_{1\_B0\_M} &= E_{1\_B1\_M} = \frac{A_\parallel}{4}, \\
        E_{0\_B1\_M} &= -\frac{A_\parallel}{4} + \frac{A_\perp}{2}\frac{\alpha}{\Delta}, \,\text{and} \\
        E_{0\_B0\_M} &= -\frac{A_\parallel}{4} - \frac{A_\perp}{2}\frac{\alpha}{\Delta},
    \end{split}
\end{equation}
where $A_\parallel$ is the component of the hyperfine interaction parallel the \DDDD{} symmetry axis, $A_\perp$ is the component perpendicular the \DDDD{} axis, $\alpha$ is the transverse strain applied to the defect, and $\Delta=\sqrt{\lambda^2 + \alpha^2}$ is the splitting due to spin-orbit coupling ($\lambda$) and strain.

We have assigned each energy level to a joint broker and memory state, as shown in schematically in Fig.~\ref{fig:intro}f.
In spite of the degeneracy of the $\ket{1\_B}$ states, this system allows direct control of the joint broker-memory system with the microwave transitions shown in blue/green (purple), corresponding to rotations of the broker (memory) qubit controlled by the memory (broker).
Driving both the blue and green transitions allows for arbitrary single qubit rotation of the broker.
By combining controlled $\pi$-rotations of the broker and memory driving all three transitions, we can a perform SWAP gate between the broker and memory.
We are therefore able to completely control the two qubit system through a combination of controlled rotations and single-qubit rotations on the broker qubit, satisfying requirements 2 and 3 in Section~\ref{sec:intro}.

Similar to the $m_S=0$ state for the \nv{}, the degeneracy of the $\ket{1\_B0\_M}$ and $\ket{1\_B1\_M}$ hyperfine states means that optical excitation does not disturb information in the memory qubit.
As illustrated in Fig.~\ref{fig:intro}f, while Hamiltonian parameters are different in the ground and excited states, the degeneracy of the $\ket{1\_B0\_M}$ and $\ket{1\_B1\_M}$ states remains unchanged, ensuring $\Delta\omega\_M$ in equation~\ref{eq:excitation_fidelity} is identically 0 when no magnetic field is applied.
We are thus able to optically excite the system when it is in either of the $\ket{1\_B}$ states without disturbing information stored in the memory qubit, satisfying requirement 1 in Section~\ref{sec:intro}.

\section{Methods \& Results}\label{sec:results}

To demonstrate this zero-field \textsuperscript{117}Sn protocol, we integrate isotopically purified \hsnv{} centers in quantum microchiplets (QMCs) onto a photonic integrated circuit (PIC; Fig.~\ref{fig:intro}e; Appendix~\ref{sec:device})~\cite{Starling2023}.
We perform all measurements in a 1.3~K ICE Oxford cryostat, dynamically addressing the optical transitions of the \hsnv{} color center through the PIC's fiber interfaces using electro-optic modulation of a continuous wave laser and applying microwave spin drive via a PIC-integrated line (Appendix~\ref{sec:setup}).

\subsection{Optical Properties}\label{sec:optical}

\begin{figure*}[ht]
    \centering
    \includegraphics[scale=0.75]{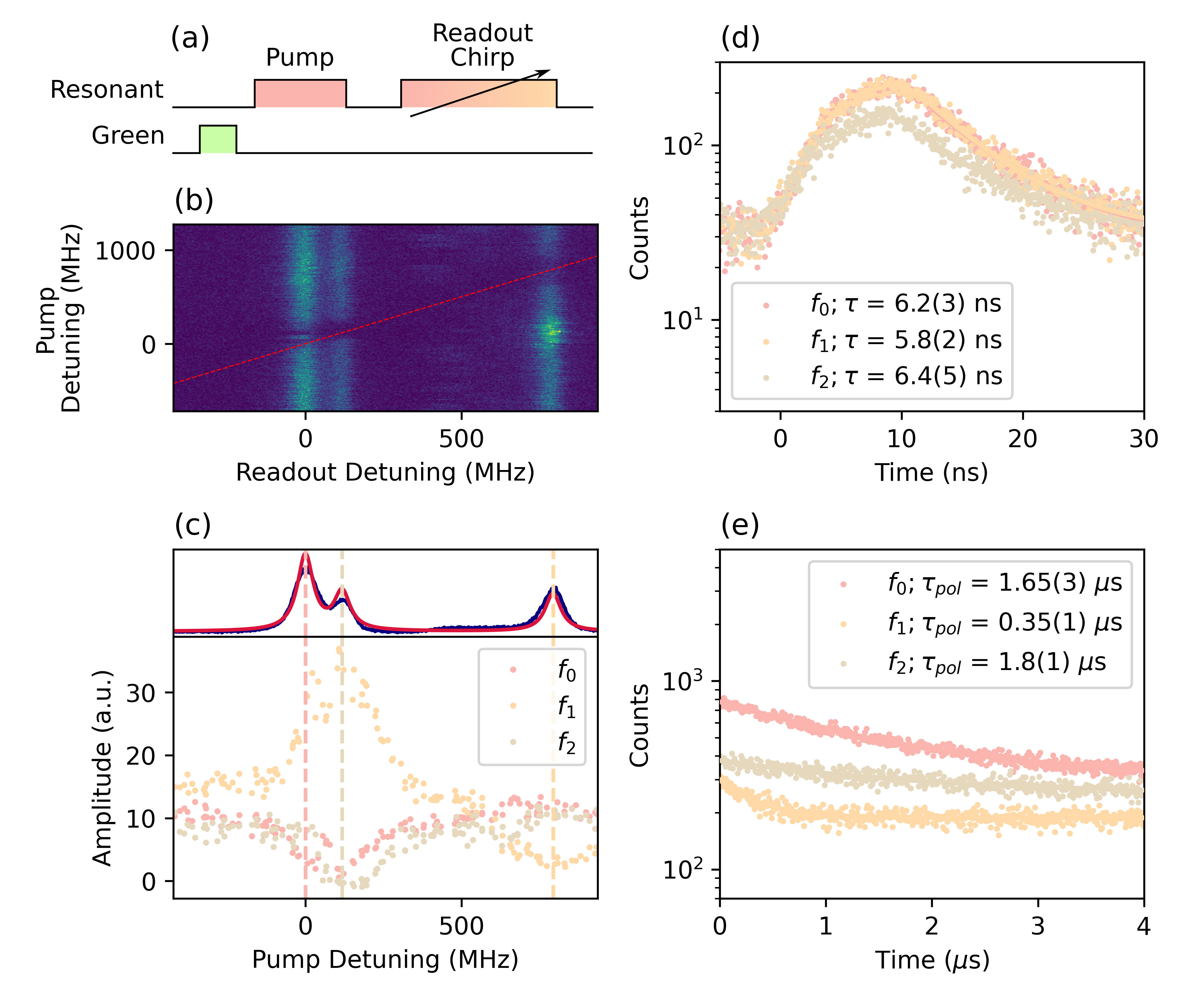}
    \caption{
    (a) Schematic of pump-probe scheme
    (b) Readout intensity as a function of readout and pump frequency.
    (c) Fitted PLE peak intensity as a function of pump frequency.
    Inset on top shows how the PLE spectrum aligns with the unperturbed PLE peak polarization.
    Maximum ground state polarization of 98.4\% occurs when pumping at 107~MHz.
    (d) Optical lifetime under resonant excitation for each transition.
    (e) Optical polarization time for each transition.
    } 
    \label{fig:init}
\end{figure*}

We first identify a well-coupled emitter with narrow linewidth by performing photoluminescence excitation (PLE) experiments in a single waveguide of the QMC.
We observe an emitter with a linewidth of 61.859(2) MHz, within a factor of 2.5 of the lifetime-limited linewidth, as shown in Fig.~\ref{fig:intro}c.

Differences in the hyperfine interaction between the ground and excited state manifolds lead to large frequency differences in the spin-conserving optical transitions (Fig.~\ref{fig:intro}b).
The eigenenergies described in equation~\ref{eq:levels} result to two sets of PLE peaks: one (corresponding to the $\ket{1\_B}$ ground states) split by magnetic field and the other (corresponding to $\ket{0_B}$) split by strain.
In our system, operating with no applied magnetic field and high strain, the $\ket{1\_B}$ transitions remains degenerate (peak at $f_0$ in Fig.~\ref{fig:intro}f), while the $\ket{0\_B}$ are separated by 675.9(4)~MHz (peaks $f_1/f_2$ in Fig.~\ref{fig:intro}f).
The large strain-induced splitting is estimated at 928~GHz in the ground state, in line with estimates of $\sim$900~GHz caused by thermal mismatch of the diamond with the substrate during cooling (see Appendix~\ref{sec:thermal_strain}). 

We demonstrate optical initialization of the hyperfine state by performing the pump-probe sequences shown in Fig.~\ref{fig:init}a, where we first apply an extended resonant 10~$\upmu$s pulse at a single frequency, followed by frequency-resolved read-out of the hyperfine state with a chirped pulse.
As shown in Fig.~\ref{fig:init}b, we can identify where resonant pumping depletes population in the resonantly pumped state and populates the other peaks.

In Fig.~\ref{fig:init}c, we plot the fitted PLE peak intensities for each pump frequency as a function of pump detuning from the $f_0$ peak frequency.
We observe depletion of each peak when the pump frequency is resonant with the peak, indicating polarization of the ground hyperfine states.
Taking the peak intensity to be directly proportional to the ground state population, we estimate the spin polarization into a given state as ratio of the corresponding peak intensity to the sum of all peak intensities.
We achieve the highest spin polarization of 98.4\% into the $\ket{0\_B0\_M}$ state by resonantly pumping near $f_2$, detuned from the $\ket{1\_B}$ peak $f_0$ by 107~MHz.

Under the standard group-theory-derived model of the group-IV color centers~\cite{Hepp2014a}, the $f_0$ transitions (from the $\ket{1\_B}$ states) should be perfectly cycling in the zero-field regime.
On the other hand, the $\ket{0\_B}$ states will have different amounts of mixing of the $\ket{\uparrow\Downarrow}$/$\ket{\downarrow\Uparrow}$ levels (anti-aligned electron $\uparrow$/$\downarrow$ and nuclear spins $\Uparrow$/$\Downarrow$) in the ground/excited state due to differences in spin-orbit coupling and strain between the two manifolds, and the state-flipping transitions between these two states should be allowed.
The theory therefore predicts that transitions $f_{1/2}$ should be optically polarizable, but driving $f_0$ should not result in any ground state polarization.
We instead observe that driving any of the transitions can optically polarize the hyperfine state.

We quantify this discrepancy by first measuring the optical lifetime of each transition under pulsed resonant excitation, as shown in Fig.~\ref{fig:init}d, yielding a lifetime of $\tau =6$~ns for each transition, in line with previous measurements for \snv{} with spin-free isotopes~\cite{Trusheim2018}.
We next measure the optical polarization time $\tau\_{pol}$ under strong optical pumping well above saturation, shown in Fig.~\ref{fig:init}e.
We use these two values to find the cyclicity $\Lambda$ from the relation~\cite{Rosenthal2024}
\begin{equation}\label{eq:cyclicity}
    \Lambda=\tau\_{pol}/2\tau.
\end{equation}
We estimate the cyclicity for transition 0, 1, and 2 at 132.44(3), 30.41(3), and 141.50(5) respectively.
By comparing the count rate at the start of the pumping pulse with the count rate expected from the 6~ns lifetime, we further estimate a total setup collection efficiency into the phonon sideband of $1.4\times10^{-4}$.

The finite cyclicity of the $f_0$ transition can be explained by adding a transverse magnetic field (see Appendix~\ref{sec:hyperfine_hamiltonian}).
In the large strain regime in which this device operates, the transverse magnetic field mixes the $\ket{1\_{B}}$ states (total electron and spin angular momentum $J=1, m_J=\pm1$) with the  $\ket{0\_B1\_M}$ state ($J\approx1, m_J=0$).
Under the conditions estimated for our device in Table~\ref{tab:parameters}, the transverse magnetic field would need to be $\sim$200~$\upmu$T to produce the cyclicity we observe (see Appendix~\ref{sec:cyclicity}), which is larger than Earth's magnetic field at our location.
Other effects beyond the standard group theory-derived Hamiltonian, such as Jahn-Teller distortion or higher-order coupling of hyperfine coefficients with strain may also explain the finite cyclicity.
Further magneto-optic and strain-dependent studies of the cyclicity are therefore needed to unambiguously determine the origin of the finite cyclicity.

\subsection{Spin Manipulation}\label{sec:spin}

\begin{figure*}[ht]
    \centering
    \includegraphics[scale=0.75]{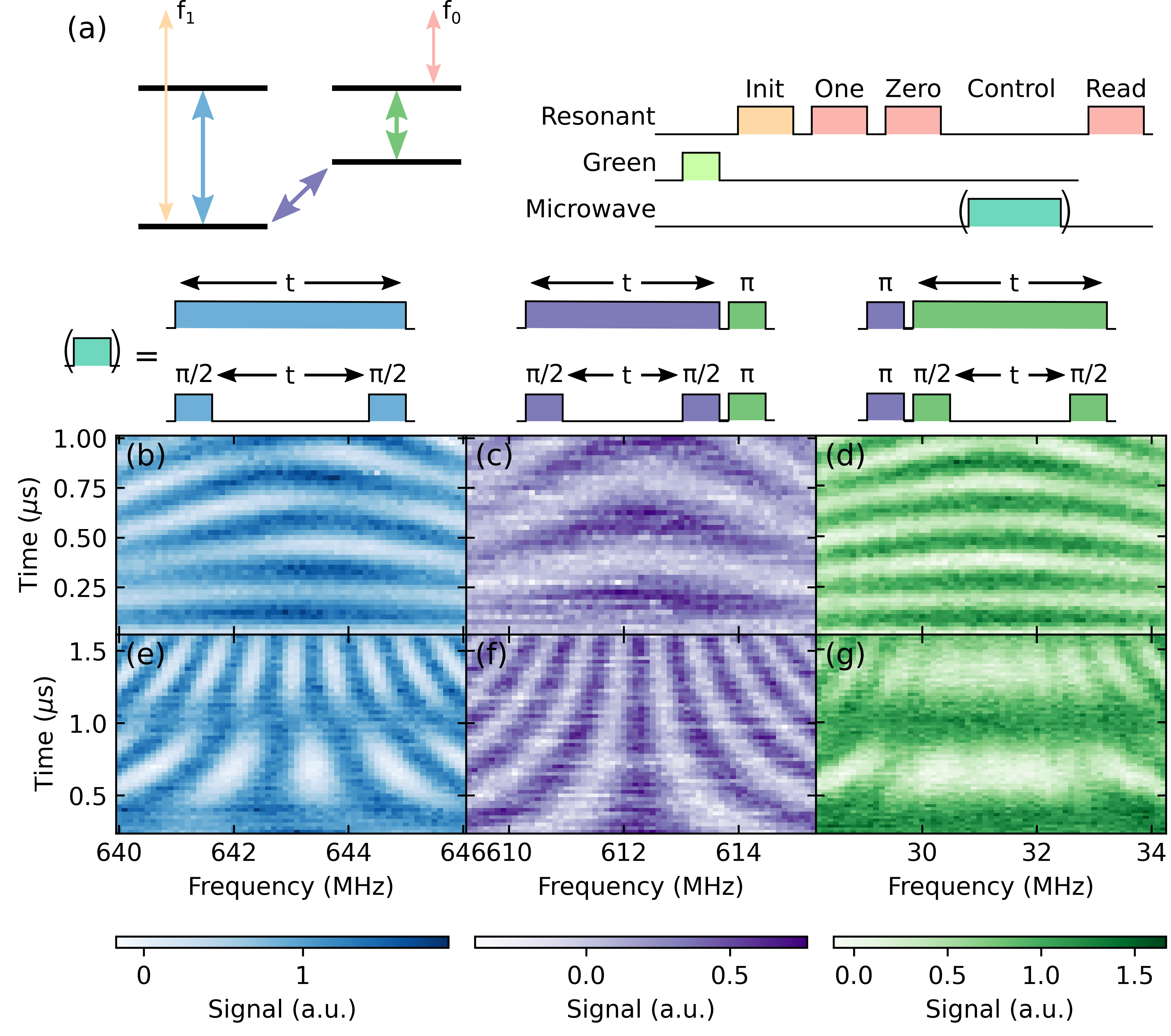}
    \caption{
    (a) We investigate the ground state manifold (left) via microwave and optical driving. 
    Our standard spin initialization sequence (right) contains the resonant and repump pulses to initialize the system while also acquiring the information necessary to normalize the results.
    For each sequence, the microwave experiment under test is nested between the initialization and readout optical pulses.
    For each transition, we show the normalized intensity for 
    (b-d) Rabi and
    (e-g) Ramsey driving.
    Notice that while the \purple{} transition (c,f) is symmetric, the other transitions (b,e and d,g) show beating and asymmetric behavior. We attribute this to residual splitting between the $1\_B$ levels (Appendix~\ref{sec:fitting}, Figs.~\ref{fig:rabi_fitting},\ref{fig:ramsey_fitting}).
    } 
    \label{fig:spin}
\end{figure*}

The transitions between the ground hyperfine states can be driven by an AC magnetic field along the defect's axis of symmetry for the $0\_B1\_M \leftrightarrow 0\_B0\_M $ transition, and perpendicular to the axis of symmetry for the other two transitions.
We use the microwave line integrated into the PIC to drive these hyperfine ground state transitions, while optically initializing into the state $\ket{0\_B0\_M}$ and reading out the $\ket{1\_B}$ state by optically addressing the $f_0$ peak.
Our optical sequence (Fig.~\ref{fig:spin}a) includes normalization measurements for the $\ket{0\_B}$ (dark) and $\ket{1\_B}$ (bright) states.
We thus directly probe the $0\_B0\_M \leftrightarrow 1\_B0\_M $ transition, which we find at 643.3~MHz (Fig.~\ref{fig:spin}b).

We find that initialization into $\ket{0\_B0\_M}$ and readout on $\ket{1\_B}$ provides the best SNR, so to characterize the remaining transitions in the spin manifold, we apply additional pre- and post- MW $\pi$-pulses to probe the $0\_B0\_M \leftrightarrow 0\_B1\_M $ and $0\_B1\_M \leftrightarrow 1\_B1\_M $ transitions which we find at 612.4 and 31.3~MHz respectively (see Fig.~\ref{fig:spin}c-d).
From these measurements, we are able to infer the hyperfine parameters $A_{\perp/\parallel}$ and strain magnitude $\alpha$ (see Appendix~\ref{sec:fitting}), which we summarize in Table~\ref{tab:parameters}.

With a roughly micrometer square cross-section and sub-micrometer distance from the diamond waveguide, the integrated microwave line permits operation on all these transitions at $>1$~MHz Rabi frequencies with a small \mbox{-8~dBm} microwave input power.
At higher microwave powers (14~dBm), we are able to drive the emitters at Rabi frequencies exceeding 20~MHz.
However, increased driving power leads to a Bloch-Seigert-like AC shift of the transition frequency due to the Rabi frequency approaching the 31~MHz frequency of the $0\_B1\_M \leftrightarrow 1\_B1\_M $ transition.
For simplicity we perform all further experiments at low power to avoid this effect.

\begin{table}
    \centering
    \begin{tabular}{c|c}
        Parameter & Value \\\hline
        $A_\perp\^{gnd}$ & 671~MHz\\ 
        $A_\parallel\^{gnd}$ & 674~MHz\\
        $\alpha\^{gnd}$ & 928~GHz \\
        $A_\perp\^{exc}$ & 464~MHz \\
        $A_\parallel\^{exc}$ & -232~MHz \\
        $\alpha\^{exc}$ & -209~GHz
    \end{tabular}

    \caption{Estimates of device \hsnv{} parameters.}
    \label{tab:parameters}
\end{table}

\subsection{Spin Fidelity}\label{sec:fidelity}

\begin{figure*}[ht]
    \centering
    \includegraphics[scale=0.75]{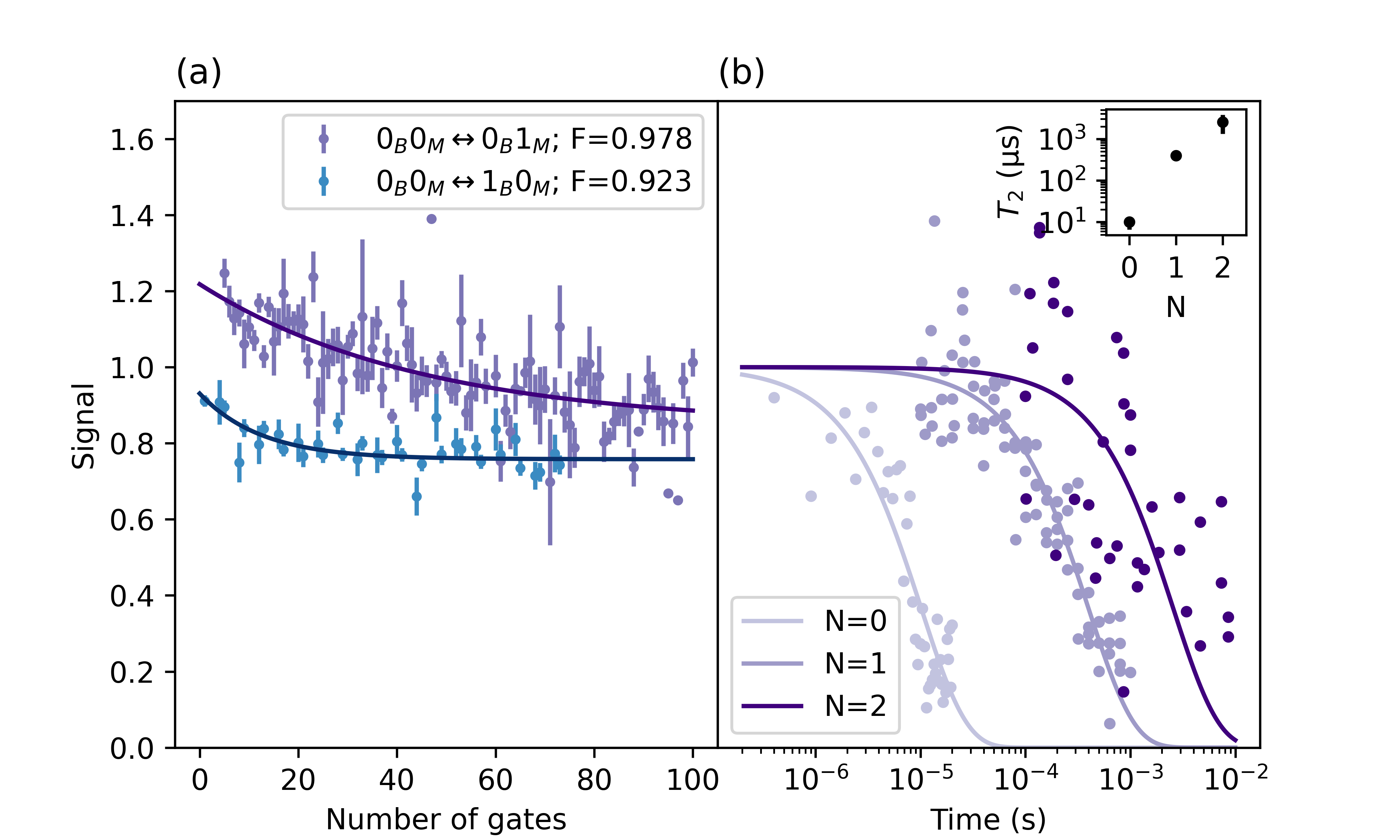}
    \caption{
    (a) Randomized benchmarking on the gate set $\left\{ R\_X( \pm\pi/2), R\_X( \pm\pi), R\_Y(\pm\pi/2), R\_Y(\pm\pi)\right\}$ for the \purple{} (purple) and \blue{} (blue) transitions showing physical gate fidelities of 97.8\% and 92.3\% respectively.
    (b) Coherence time measurements with N=0, 1, and 2 decoupling XY pulses on the \purple{} transition. Inset shows extracted $T_2$ coherence times as a function of the number of pulses.
    }
    \label{fig:node}
\end{figure*}

We next characterize the coherence of the hyperfine levels by performing Ramsey sequences on each transition, shown in Fig.~\ref{fig:spin}e-g.
By simulating the full Hamiltonian, as shown in Appendix~\ref{sec:fitting}, we extract $\ttwostar$ of 49.2(3)~$\upmu$s for the \blue{} transition (blue), 26.4(2)~$\upmu$s for the \purple{} transition (purple), and 37.9(2)~$\upmu$s for the 
\green{} transition (green).

We note that while the \purple{} transition exhibits the normal Ramsey fringe pattern, the other two transitions show additional interference bands at a 1.4 MHz frequency.
We attribute this to the imperfect degeneracy of the $\ket{1\_B0\_M}$ and $\ket{1\_B1\_M}$ states.
If the states are perfectly degenerate, microwave selection rules dictate that the double state-flipping transitions $0\_B1\_M\leftrightarrow 1\_B0\_M$ and $0\_B0\_M\leftrightarrow 1\_B1\_M$ are not allowed.
However, in the presence of a small residual coupling between the two $\ket{1\_B}$ states, these selection rules are broken, and free precession produces the interference pattern seen in Fig.~\ref{fig:spin}e,g.
The coupling is well explained by adding a DC magnetic field with a total amplitude of 223~$\upmu$T, again larger than Earth's magnetic field, but in close agreement with the value predicted from the cyclicity (see Appendix~\ref{sec:fitting}).
Jahn-Teller distortion and strain-hyperfine coupling mentioned previously, which break the degeneracy of the $A_{xx}$ and $A_{yy}$ components of the hyperfine coupling, may also explain the non-degeneracy.


In order to show that the \hsnv{} ground state can be controlled with high fidelity under the proposed zero-field regime, we next demonstrate high fidelity operation of $\pi$ and $\pi/2$ rotations on the \purple{} (purple) transition using gate times calibrated from the Rabi driving in Section~\ref{sec:spin}.
By shifting the phase of the microwave drive between pulses we are able to perform rotations around arbitrary axes in the X/Y plane.
We perform randomized benchmarking by applying a series of gates of length $N$ randomly selected from the set $\left\{ R\_X( \pm\pi/2), R\_X( \pm\pi), R\_Y(\pm\pi/2), R\_Y(\pm\pi)\right\}$, where the last gate is chosen to bring the qubit into the bright $\ket{1\_B}$ state (in combination with a final \green{} $\pi$-pulse).
Measuring the signal with the same readout and initialization scheme as in Fig.~\ref{fig:spin}a, we extract an estimate of the physical gate fidelity of 97.8\% (purple in Fig.~\ref{fig:node}a).
By operating at a reduced -34 dBm input power, we are able to resolve the near-degenerate $\ket{1\_B}$ states, and via a similar randomized benchmarking sequence selectively address the \blue{} (blue) transition with a 92.3\% fidelity (blue in Fig.~\ref{fig:node}a).
Adjusting for the fact that the Clifford gate set should include the additional 5 gates $\left\{ I, R\_Z( \pm\pi/2), R\_Z( \pm\pi)  \right\}$, which can be performed by adjusting the phases of the physical gate pulses, this results in a single-qubit Clifford gate fidelity of 98.6\% for the memory and 95.2\% for the broker respectively.

Finally, using the center frequency from the 2D Ramsey experiments, we perform XY decoupling pulse sequences with 1 and 2 decoupling $\pi$-pulses on the \purple{} transition, extending the coherence time to 400(75)~$\upmu$s with one pulse and 2.5(1.2)~ms with two.
The measurement is limited only by the long experiment times required due the relatively low collection efficiency of the system, and we expect longer coherence times should be possible with more decoupling pulses. 

With a comparable number of decoupling pulses, similar measurements with the bare electron \snv{} spin only have coherence times ranging from 100-400~$\upmu$s~\cite{Guo2023,Rosenthal2023,Karapatzakis2024}, an order of magnitude less than the 2.5~ms coherence time we measure.
This can be explained using the theory of coherence for group-IV color centers~\cite{Harris2024}.
As detailed in Appendix~\ref{sec:coherence}, while the broker qubit subspace is susceptible to decoherence from first-order phonon processes similar to the bare electron in \snv{}, the memory qubit is immune to phonon-mediated dephasing.
Further, the \purple{} transition we measure here is insensitive to magnetic field to first order, decoupling it from spin-bath noise.
We expect the dominant source of noise on this transition to be strain noise, possibly from charge re-arrangement in the vicinity of the color center.
Thus, in spite of the strong hyperfine coupling to the electron spin, the memory qubit subspace maintains good ``nuclear-like'' coherence properties.

\section{Conclusion \& Discussion}\label{sec:conclusion}

We have presented a zero magnetic field bias scheme for \hsnv{} quantum network nodes, in which the level structure contains broker and memory qubit subspaces.
Under this scheme, the broker qubit can be optically excited without causing undesired decoherence of the local memory qubit.
This is made possible by the large hyperfine coupling of \hsnv{} compared to other group-IV color centers, which allows direct optical access of the hyperfine levels at zero magnetic field.
In spite of the large coupling, the memory qubit is not affected by phonon-mediated dephasing.

We have integrated a \hsnv{} color center into a PIC device and operate it under zero magnetic field, showing:
\begin{enumerate}
    \item ground state optical polarization of 98.4\%,
    \item Clifford gate fidelity of the memory and broker qubits of 98.6\% and 95.2\% respectively with gate speeds in excess of 1~MHz,
    \item coherence times of the memory transition extending to 2.5~ms with only 2 decoupling pulses.
\end{enumerate}
This demonstrates that high fidelity operation of the \hsnv{} is possible under the zero-field scheme we propose.
Our device is primarily limited by the low collection efficiency into the phonon sideband of $1.4\times10^{-4}$, however this is not a fundamental limitation of the system.
In comparison, recent state-of-the-art results have shown collection efficiency of \hsnv{} into the zero phonon line of as high as $1.7\times10^{-2}$~\cite{Parker2023}.
Spin-photon fidelity can further be improved by integrating the \hsnv{} with optical cavities in the high cooperativity regime, which has also been demonstrated in previous work with group-IV color centers~\cite{Stas2022,Knaut2024}, including \snv{}~\cite{Rugar2021,Kuruma2021,Herrmann2023,Chen2024}.
In this regime, single-shot readout and heralded spin-spin entanglement should be achievable.

We have also observed two significant deviations of the \hsnv{} behavior from the predictions of the standard theory of group-IV color centers at zero magnetic field, namely the finite optical cyclicity of the $f_0$ transition, and the non-degeneracy of the $m_J=1$ states.
Both effects can be explained by a weak residual magnetic field, however the required field strength is an order of magnitude stronger than the strongest magnetic field we expect in our setup, Earth's magnetic field.
Further magneto-optic, strain-optic, and theoretical studies are needed to confirm or eliminate the residual magnetic field explanation for these effects (see Appendix~\ref{sec:fitting}).
One alternative explanation is coupling of the hyperfine constant to strong bulk or Jahn-Teller strain present in the defect.
As both distortions break the degeneracy of the $A_{xx}$ and $A_{yy}$ hyperfine components, they could explain the non-degeneracy and finite cyclicity.
Understanding of the transition cyclicity can also help improve single-shot readout fidelity by increasing the cyclicity of the $f_0$ readout transition; for example it may be possible to increase cyclicity with lower strain or a small magnetic field along the \DDDD{}-axis of the defect.
More strain- and temperature-dependent measurements and theoretical study are also needed to quantify the nuclear spin-orbit coefficient, $\upsilon$, which has an important effect on the coherence of the broker qubit (see Appendix~\ref{sec:coherence}).
Further, the level structure presented here is not unique to \hsnv{}: other diamond color centers such as \textsuperscript{73}GeV\textsuperscript{--}~\cite{Adambukulam2024}, \textsuperscript{207}PbV\textsuperscript{--}~\cite{Trusheim2019,Wang2021a}, and \textsuperscript{61}NiV\textsuperscript{--}~\cite{Gali2021,Morris2024} may also be amenable to similar zero-field operation.
Color centers in other materials, such as Ytterbium ions in YVO~\cite{Kindem2020}, and V\textsuperscript{4+} in SiC~\cite{Wolfowicz2020} may also have similar level structure with optically resolvable hyperfine levels at zero field which could benefit from operating under our protocol.

The \hsnv{} also offers novel opportunities for interacting with additional \cspin{} or other proximal spins to further increase the number of available memory qubits.
Current protocols for \cspin{} control using \nv{}~\cite{Abobeih2022,Bradley2019} or the bare \snv{}~\cite{Debroux2021,Beukers2024a} must contend with the fact that the hyperfine interaction with the electron is always on.
When the broker qubit in \hsnv{} is in the $\ket{0\_B}$ state, the memory qubit is insensitive to magnetic field to first order, and is therefore decoupled from small hyperfine interactions with other nuclei.
By driving one or both of the \blue{} or \green{} transitions, information stored in the \hsnv{} memory qubit can be conditionally or unconditionally coupled to local nuclei.
Further, if the $f_1$ optical transition is addressed, the proximal spins should remain decoupled in the excited state, meaning the \cspin{} qubits are insensitive to optical excitation of the \hsnv{} ``memory qubit'', just as the memory qubit is insensitive to excitation of the broker within \hsnv{}.

The proposal presented in this paper resolves well-known challenges associated with \textsuperscript{29}Si or \cspin{} quantum network memories in group-IV color centers~\cite{Stas2022,Knaut2024,Beukers2024a}.
The \hsnv{} system provides a local memory which is deterministically-included, immune from dephasing caused by optical readout or entanglement attempts, and where both the memory and qubit gates can be driven at high speed.
The zero-field and moderate-temperature operating conditions of our protocol allow deployment without expensive superconducting magnets and in systems with orders of magnitude greater cooling power 
than dilution refrigerators.
Indeed, our high fidelity control of a \hsnv{}-based device demonstrates the viability of zero-field operation and paves the way for the use of this protocol in quantum networks, local qubit control, and the fundamental understanding of color center physics.

\begin{acknowledgements}
This work was supported in part by the STC Center for Integrated Quantum Materials (CIQM) NSF Grant No. DMR-1231319, the National Science Foundation (NSF) Engineering Research Center for Quantum Networks (CQN) awarded under cooperative agreement number 1941583, and the MITRE Moonshot Program.
Additional support provided by European Union’s Horizon 2020 research and innovation program (grant agreement No.896401).
I.C. and O.M. acknowledge support from the National Defense Science and Engineering Graduate Fellowship (NDSEG).
M.C. acknowledges support from the MIT Claude E. Shannon Award.

Distribution Statement A. Approved for public release. Distribution is unlimited. This material is based upon work supported by the National Reconnaissance Office and the Under Secretary of Defense for Research and Engineering under Air Force Contract No. FA8702-15-D-0001. Any opinions, findings, conclusions or recommendations expressed in this material are those of the authors and do not necessarily reflect the views of the National Reconnaissance Office or the Under Secretary of Defense for Research and Engineering. © 2025 Massachusetts Institute of Technology. Delivered to the U.S. Government with Unlimited Rights, as defined in DFARS Part 252.227-7013 or 7014 (Feb 2014). Notwithstanding any copyright notice, U.S. Government rights in this work are defined by DFARS 252.227-7013 or DFARS 252.227-7014 as detailed above. Use of this work other than as specifically authorized by the U.S. Government may violate any copyrights that exist in this work.

\end{acknowledgements}

\bibliography{MSc}

\clearpage
\appendix

\section{Optical Insensitivity}\label{sec:optical_insensitivity}
In color centers, the broker qubit is normally the electronic spin coupled to an optical transition. 
If the electron spin states interact with a nearby nuclear spin, the hyperfine interaction will provide an additional memory degree of freedom, splitting the levels with a frequency $\omega\_M\^{gnd/exc}$ depending on whether the electron is in the ground or optically excited state.
Since the electron's distribution around the nucleus is not the same in the ground and excited states, in general $\omega\_M\^{gnd}\neq \omega\_M\^{exc}$.
If a quantum state is stored within these hyperfine levels, during optical excitation, the state will accumulate a phase
\begin{equation}\label{eq:excitation_effect}
    \ket{\psi} = \alpha\ket{0} + \beta\ket{1} \rightarrow \alpha\ket{0} + e^{i(\omega\_M\^{exc} - \omega\_M\^{gnd} )t\_{exc}}\beta\ket{1}
\end{equation}
where $t\_{exc}$  is the time spent in the excited state.

The time $t\_{exc}$ is generally not known, as the optical collection efficiency $\eta\ll 1$ in realistic networking setups.
If the time in the excited state is sampled from the probability distribution $p(t\_{exc})$ after the excitation, if the memory is in the initial state $\rho$, the new state will be
\begin{equation}
    \rho \rightarrow \int_0^\infty U(t\_{exc})\rho U^\dagger(t\_{exc}) \, p(t\_{exc}) dt\_{exc}
\end{equation}
where $U(t\_{exc})$ produces the transformation in equation~\ref{eq:excitation_effect}.

If the emitter is excited by an optical $\pi$ pulse, and is then allowed to decay with lifetime $\tau$, then $p(t\_{exc})=\tau^{-1}\exp(-t\_{exc}/\tau)$.
If the initial memory state is a bell state, $\ket{\psi} = \left(\ket{0}+\ket{1}\right)/\sqrt{2}$, then the fidelity after $n$ optical excitations with mean lifetime $\tau$ (up to a correctable phase) will be
\begin{equation}
    F = \frac{1}{2}\left( 1 + (1 + \Delta\omega\_M^2\tau^2)^{-n/2} \right).
\end{equation}



\section{\hsnv{} Ground State}\label{sec:hyperfine_hamiltonian}
\subsection{Hamiltonian}
As outlined in~\cite{Harris2023}, the level structure of the \hsnv{} is described through the interplay of the orbital, electron spin, and nuclear spin degrees of freedom.
In the absence of magnetic field, the system's Hamiltonian is written
\begin{equation}\label{eq:hamiltonian}
    \hat{H} = \hat{H}\_{SOC} + \hat{H}\_{Egx} + \hat{H}\_{Egy} + \hat{H}\_{HF} + \hat{H}\_{IOC}
\end{equation}
where $\hat{H}\_{SOC/IOC}$ are the electron/nuclear spin-orbit coupling, $\hat{H}\_{Egx/y}$ is the magnitude of the transverse symmetry-breaking strain applied on the defect, and $\hat{H}\_{HF}$ is the hyperfine coupling.
In a basis spanned by the orbital, electron, and nuclear states $\ket{e_{g\pm}}\_{L} \otimes \ket{\ud}\_{S} \otimes \ket{\UD}\_{I}$, these terms are inferred from group theory to be~\cite{Hepp2014a}
\begin{equation}
    \begin{split}
        \hat{H}\_{SOC} &= \frac{1}{2}\lambda\,\hat\sigma_z\^{L}\hat\sigma_z\^{S}, \\
        \hat{H}\_{IOC} &= \frac{1}{2}\upsilon\,\hat\sigma_z\^{L}\hat\sigma_z\^{I}, \\
        \hat{H}\_{Egx/y} &= -\alpha\_{Egx/y}\hat\sigma_{x/y}\^{L}, \\
        \hat{H}\_{HF} &= \frac{1}{4}A_\perp \left( \hat\sigma_x\^{S}\hat\sigma_x\^{I} + \hat\sigma_y\^{S}\hat\sigma_y\^{I} \right) + \frac{1}{4}A_\parallel\hat\sigma_z\^{S}\hat\sigma_z\^{I}
    \end{split}
\end{equation}
where $\hat\sigma_{i}\^{j}$ are the Pauli matrices ($i\in\{x,y,z\}$ in the spatial basis shown in Fig.~\ref{fig:intro}d) applied to the orbital, electron, and nuclear degrees of freedom for $j=L,S,I$ respectively, $\lambda/\upsilon$ are the strengths of the electron/nuclear spin-orbit coupling, $\alpha\_{Egx/y}$ is the magnitude of the applied transverse strain, and $A_{\perp/\parallel}$ is the strength of the hyperfine coupling.

\subsection{Energy Levels}
The full eigenenergies of equation~\ref{eq:hamiltonian} are difficult to work with analytically.
We can gain insight into the level structure using the fact that $A_{\perp/\parallel}, \upsilon \ll \Delta$, where $\Delta$ is the spin-orbit-strain splitting $\Delta=\sqrt{\lambda^2 + \alpha\_{Egx}^2 + \alpha\_{Egy}^2}$.
This allows us to expand to first order in $A_{\perp/\parallel}/\Delta$, yielding a set of energies which we label by the angular momentum of the electron and nuclear spins $J$, and the alignment of the total angular momentum along the defect's symmetry axis $m_J$:
\begin{equation}\label{eq:energies_1st_order}
    \begin{split}
        E_{J=1}^\pm =& \frac{A_\parallel}{4} \pm \frac{\Delta}{2}, \\
        E_{m_J=0'}^\pm =& -\frac{A_\parallel}{4} + \frac{A_\perp}{2}\frac{\alpha}{\Delta} \pm \frac{\Delta}{2},  \\
        E_{m_J=0''}^\pm =& -\frac{A_\parallel}{4} - \frac{A_\perp}{2}\frac{\alpha}{\Delta} \pm  \frac{\Delta}{2}.
    \end{split}
\end{equation}

As with the bare group-IV color centers, the ground state is split by strain and electron spin-orbit coupling into two sets of energy levels in upper and lower branches separated by $\sim\Delta$, with upper/lower branch corresponding to the $\pm$ condition in equation~\ref{eq:energies_1st_order}.

Within each of the $\pm$ branch, there are two degenerate energy levels labeled $E_{J=1}$, corresponding to states where the electron and nuclear spin are aligned $\ket{\uparrow\Uparrow}/\ket{\downarrow\Downarrow}$ with total electron and nuclear spin angular momentum $J=1$.
The other two energy levels $E_{m_J=0'/0''}$ contain linear combinations of states where the electron nuclear spin are anti-aligned $\ket{\uparrow\Downarrow}/\ket{\downarrow\Uparrow}$, with the component of the total angular momentum of the electron and nuclear spin along the defect axis $m_J=0$.
At zero strain, the $E_{m_J=0'/0''}$ levels are degenerate, and split from the $E_{J=1}$ by $A_\parallel/2$.
As strain increases, the $E_{J=1}$ levels remain unchanged, while the $E_{m_J=0'/0''}$ levels become non-degenerate, separating into a triplet and singlet state separated by $A_\perp$ in the limit of large strain $\alpha\gg\lambda$.

These four levels can store two qubits of information, and we can assign which eigenstates correspond to which qubit states.
It is clear from equation~\ref{eq:energies_1st_order} that the degeneracy of the $J=1$ levels is not affected by the hyperfine parameters or strain, which may both be different between the ground and optically excited state.
We can therefore define the broker qubit to be in the state $\ket{1\_B}$ when the system is in either of the $J=1$ states, and  $\ket{0\_B}$ when in the $m_J=0$ states.
Conversely, the memory qubit state is determined by which $m_J=0$ state or which $J=1$ state the system is in.
As outlined in the main text, this choice of qubit basis has the attractive property that, so long as we only optically address the $\ket{1\_B}$ spin-conserving transition (labeled $f_0$ Fig.~\ref{fig:intro}f), we optically address the broker qubit without disturbing the memory qubit.

Since $\Delta\gg k_BT$, we can neglect the upper branch for the purpose of spin driving, as the system will overwhelmingly reside in the lower branch.
Taking only the negative conditions in equation~\ref{eq:energies_1st_order}, ignoring the constant $-\Delta/2$ offset, and re-labeling to the corresponding qubit states recovers equation~\ref{eq:levels} in the main text.

\subsection{Magnetic Field}
The addition of a non-zero magnetic field breaks the degeneracy of the $J=1$ levels, and causes hybridization of the two $m_J=0$ levels.
This is modeled by adding to the Hamiltonian in equation~\ref{eq:hamiltonian} terms for Zeeman coupling of the electron spin, nuclear spin, and orbital degrees of freedom
\begin{equation}\label{eq:magnetic}
    \begin{split}
        \hat{H}\_{Z} &= g\_e\mu\_{B} \boldsymbol{B} \cdot \hat{\boldsymbol{S}} \\
         &= \frac{1}{2}g\_e\mu\_{B} \left( B_x \sigma_x\^{S} + B_y \sigma_y\^{S} + B_z \sigma_z\^{S} \right), \\
        \hat{H}\_I &= g\_I\mu\_N\boldsymbol{B} \cdot \hat{\boldsymbol{I}} \\
        &=  \frac{1}{2}g_I\mu_N \left( B_x \sigma_x\^{I} + B_y \sigma_y\^{I} + B_z \sigma_z\^{I} \right), \\
        \hat{H}\_{L} &= \frac{1}{2}qg\_e\mu\_{B} B_z \sigma_z^\mathrm{L}
    \end{split}
\end{equation}
where $g\_{e/I}$ is the electron/nuclear gyromagnetic ratio, $\mu\_{B/N}$ is the Bohr/nuclear magneton, and q is the effective orbital gyromagnetic ratio quenching factor~\cite{Hepp2014a}.

\section{Device Preparation}\label{sec:device}
Photonic integrated circuit fabrication begins with a silicon wafer. On top of an initial 4.7~$\upmu$m oxide cladding, the microwave electrodes are fabricated using 1~$\upmu$m aluminum. These are cladded with oxide and CMP controls the thickness above the aluminum to be roughly 600~nm. Then, the photonic circuitry is fabricated from 100~nm silicon nitride.
In areas where we will remove the final 4~$\upmu$m photonic cladding, we add an aluminum etch mask.
This acts as an etch stop when etching through this thick cladding.
The etch stop is wet-etched away to expose the silicon nitride for integration with diamond.
The thick cladding is essential to get good mode matching to butt-coupled optical fiber.
An additional step exposes aluminum pads and we use a dry etch to expose and polish the optical facets.
Narrow dicing in the optical facet trenches combined with stealth backside dicing separates our 5 $\times$ 4~mm chips.

To prepare the diamond, we start by implanting a polished single-crystalline seed with Sn ions (350 keV, $1\times10^{11}$ions/cm$^2$) and annealing in a staggered ramp to form SnV centers with minimized concentrations of deleterious defects.
Nanobeam chiplets in sets of eight waveguides are fabricated with a rectangular undercut process and treated with piranha acid cleaning.

We then use a tungsten probe to break a chiplet away from the parent, transferring it onto a PDMS stamp.
The stamp is transparent, so we image through the stamp to align the waveguides to the PIC.
Finally, we write polymer blocks (UpBrix) using two-photon polymerization (UpNano) over the ends of the diamond waveguides to lock the chiplet in place and improve coupling efficiency between the diamond and the PIC.

The chip is mounted upon a copper mount using a combination of thermal grease (Apezion N) and varnish (GE). The chip pads are wirebonded to a PCB with SMP connectors for RF. The optical facet remains exposed on one side of the assembly.

\section{Experimental Setup}\label{sec:setup}

\begin{figure*}[ht]
\centering
\includegraphics[scale=0.75]{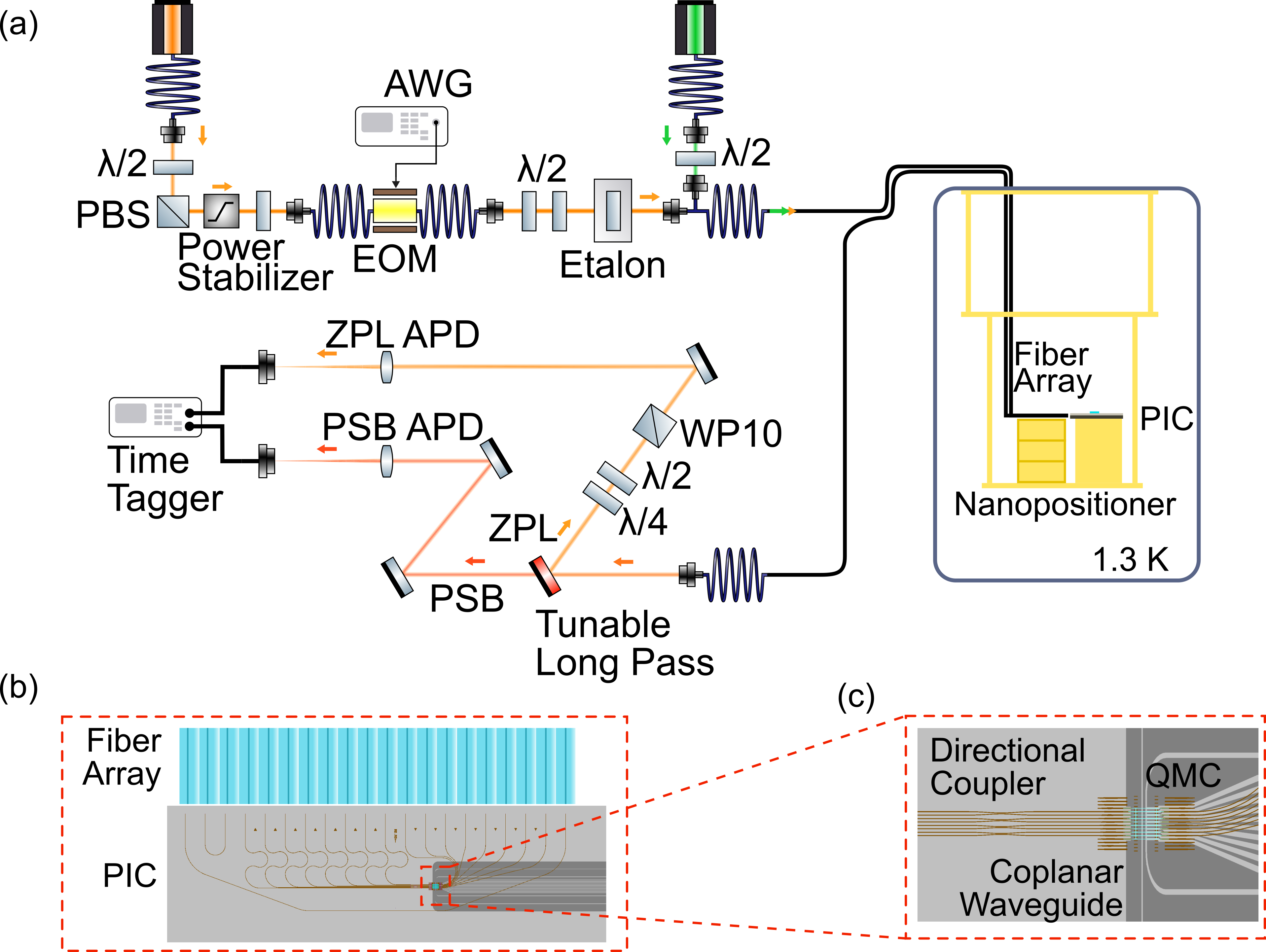}
\caption{
(a) Diagram of the optical setup and cryostat.
(b) Alignment of the fiber array with the photonic integrated circuit
(b) Zoomed in diagram of the diamond QMC integrated with coplanar waveguides and directional couplers.
}
\label{fig:hyperfine_device}
\end{figure*}

We perform all measurements in a 1.3~K ICE Oxford cryostat.
In order to access the edge-coupled waveguides in the PIC, we mount a fiber array on a piezo stack (Attocube) next to the PIC, giving us access to each channel of the PIC. 
Two channels are dedicated to an on-chip loopback, which in combination with Attocube positioner, allows the fiber to maintain alignment with the chip during cool down to 1.3~K.
To account for thermal drift in the direction of the fiber ($x$), we measure the beamwaist of scans in the other ($y$ and $z$) directions and automatically move toward and away from the chip to maintain a waist corresponding to roughly 100~$\upmu$m distance.
The rest of the channels lead to the QMC waveguides, allowing optical access, as shown in Fig.~\ref{fig:hyperfine_device}b.

We address the optical transitions of the \hsnv{} color centers using electro-optic modulator (EOM) for fast frequency control of a laser, with the setup shown in Fig.~\ref{fig:hyperfine_device}a.
Modulation with an EOM allows fast generation of sidebands around the input laser frequency $f_c$ to define resonant optical pulses.
The excitation path allows for polarization control of the EOM-modulated laser, and combines the 620~nm resonant laser with a 515~nm off-resonant laser (H\"{u}bner Cobolt) for off-resonant charge state re-pump.
The combined laser light is passed to the PIC via the fiber array, where it encounters an on-chip passive 80:20 waveguide directional coupler, before going to the \hsnv{} QMC (Fig.~\ref{fig:hyperfine_device}c).
The 80:20 beamsplitter allows us to monitor fluorescence from a neighboring waveguide channel, collecting 80\% of the fluorescence, while only 20\% of the reflected laser light is allowed through, suppressing unwanted counts from reflected red laser.
The collected light is channeled back through the fiber array, and separated out into zero-phonon-line (ZPL) emission at 620 nm, and phonon sideband (PSB) coming from phonon-assisted optical transitions.
In order to further isolate the resonant red laser, we monitor PSB fluorescence using an avalanche photodiode (APD; Excelitas) single photon detector.
Photon counts are monitored using a timetagger (PicoQuant HydraHarp 400) to give precise timing information.
The entire system is driven by an arbitrary waveform generator (Tektronix AWG70002B), which provides the RF signals for the microwave and EOM modulation, as well as TTL synchronization signals for the timetagger and green laser.


\section{Thermal Expansion Strain}\label{sec:thermal_strain}
\begin{figure*}
    \centering
    \includegraphics[width=\linewidth]{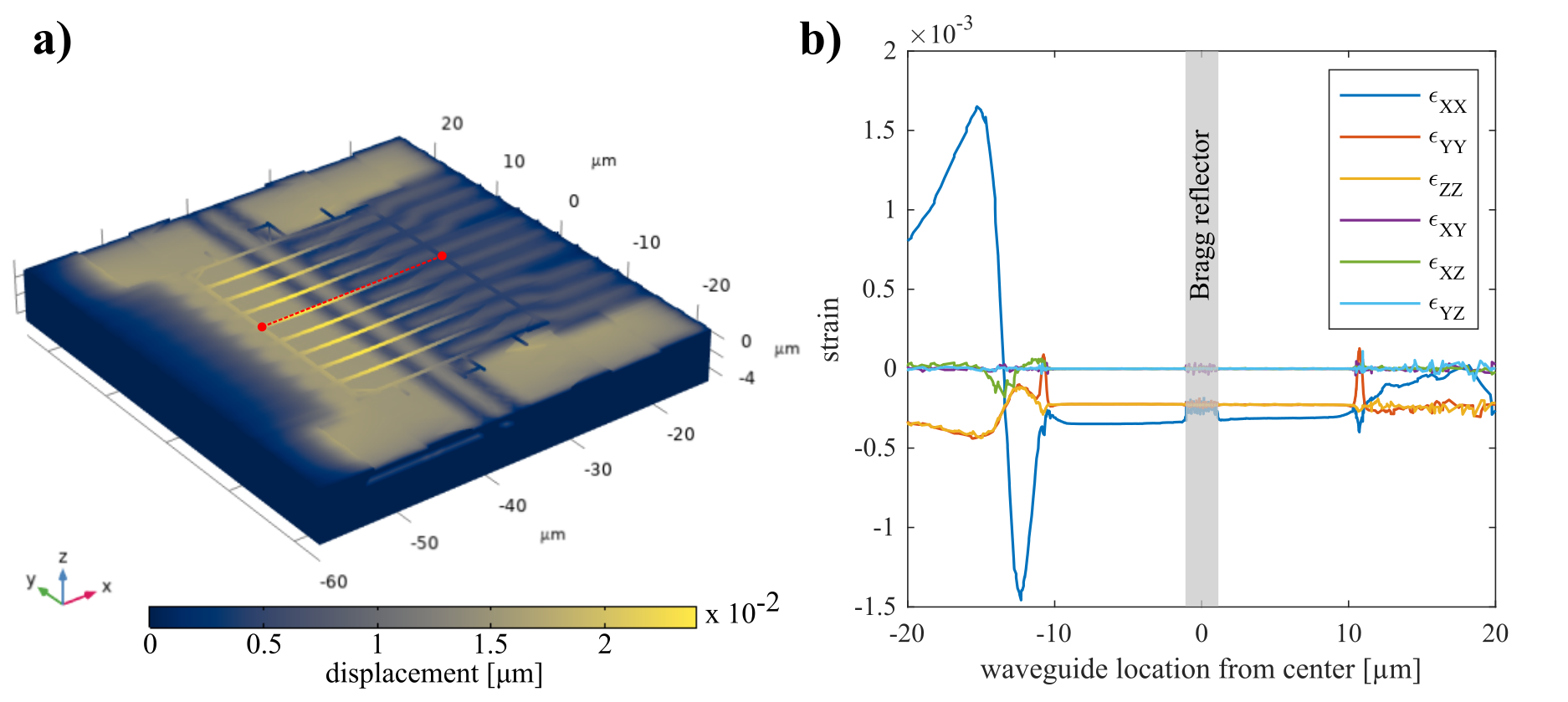}
    \caption{Strain due to thermal contraction. (a) displacement due to thermal contraction of the QMC and photonic integrated circuit. (b) Thermal contraction-induced strain tensor elements along the center of the diamond waveguide at the implantation depth of the emitter (~70 nm).}
    \label{fig:thermal_contraction}
\end{figure*}

After sample preparation at room temperature, the QMC undergoes thermal contraction upon cooling to a 1.3 K environment. We model the thermal contraction using FEM simulation in COMSOL to extract the strain in the central diamond waveguide. The thermal contraction and resulting strain along the diamond waveguide are shown in Fig. \ref{fig:thermal_contraction}. We find that the thermal strain results in a predicted $\alpha^{gnd}$ of around 900 GHz, consistent with the 928 GHz strain-induced splitting in the main text.

\section{Optical Cyclicity}\label{sec:cyclicity}
\begin{figure}[ht]\centering
\includegraphics[]{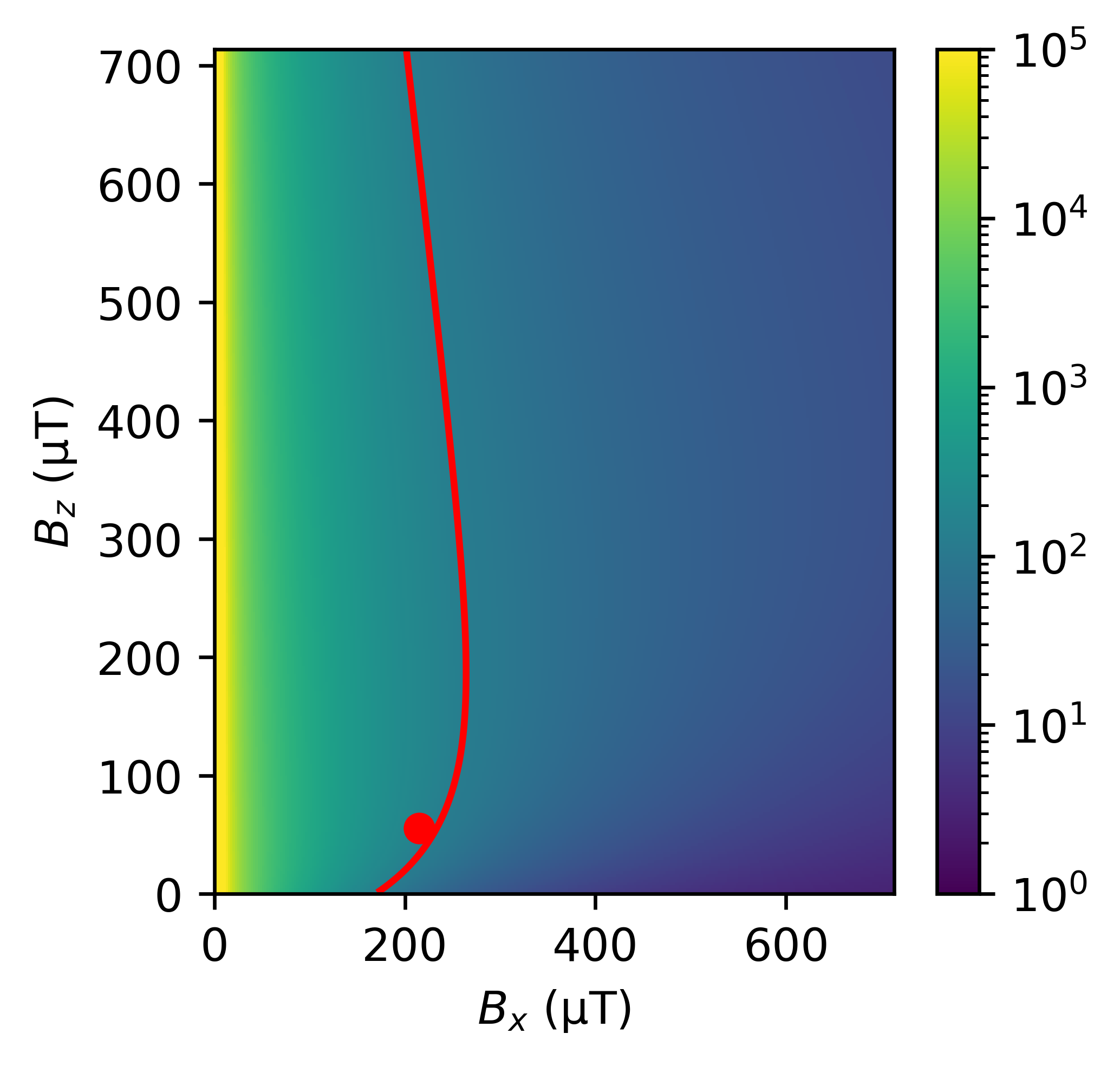}
\caption{\label{fig:cyclicity}
Optical cyclicity of the $f_0$ transition under the bias conditions shown in Table~\ref{tab:parameters} with the addition of a residual magnetic field.
The red contour line shows fields where the cyclicity values matches the experiment.
Red dot indicates magnetic field calculated from the ground state Hamiltonian.}
\end{figure}

We calculate the optical cyclicity using the transition matrix elements derived from group theory for group-IV color centers~\cite{Hepp2014a}.
Calculating the eigenstates in the ground and excited state using the parameters in Table~\ref{tab:parameters} with the addition of a small magnetic field in equation~\ref{eq:magnetic}, we calculate the cyclicity as
\begin{equation}
    \Lambda_i = \max_j\left( \frac{ \left| \melement{\psi_i\^{exc}}{\sum_k\hat{p}_k}{\psi_j\^{gnd}} \right|^2}{\sum_m\left| \melement{\psi_i\^{exc}}{\sum_k\hat{p}_k}{\psi_m\^{gnd}} \right|^2 } \right).
\end{equation}
This is plotted in Fig.~\ref{fig:cyclicity} for the $f_0$ transition as a function of axial and transverse magnetic field.
To match the observed cyclicity (red contour), a transverse field in excess of 200~$\upmu$T is required.
This is in relatively close agreement with the fit of the magnetic field perturbing the $\ket{1\_B}$ ground states (red point).

\section{Fitting Hamiltonian from Rabi and Ramsey Experiments}\label{sec:fitting}
\begin{figure*}[ht]
    \centering
\includegraphics[]{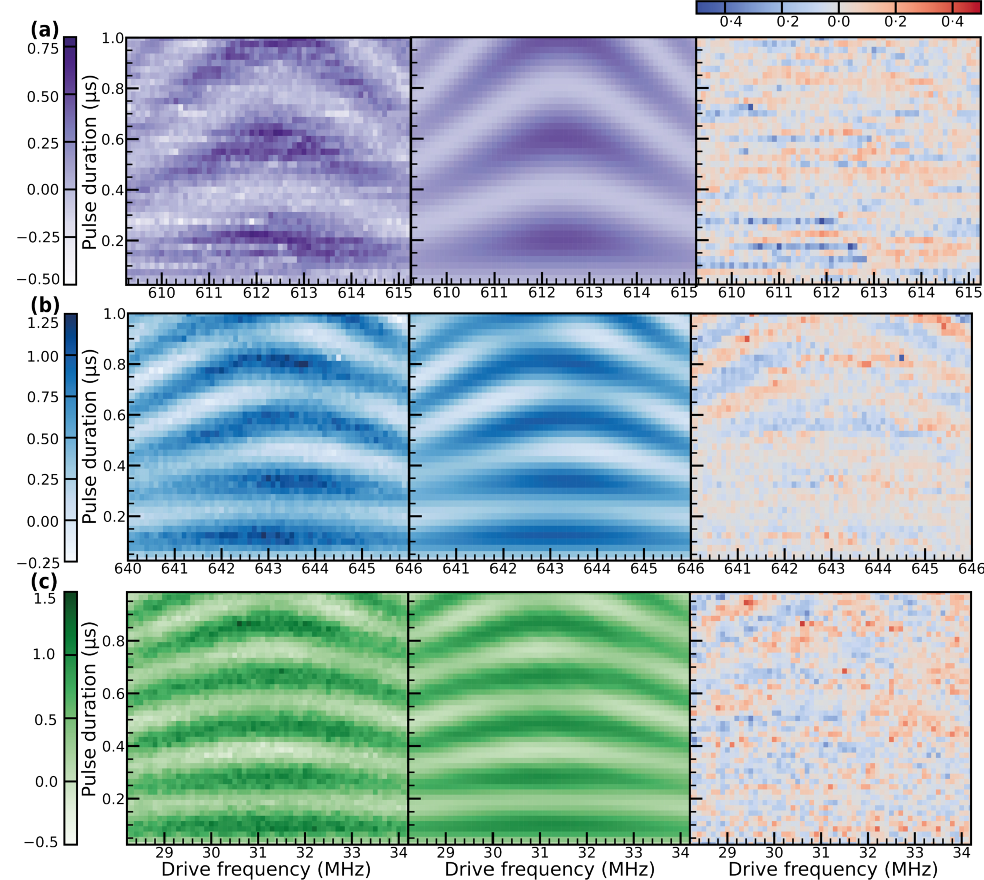}
\caption{\label{fig:rabi_fitting}Results from the combined data fit to Rabi and Ramsey oscillations from transitions (a) \purple{}, (b) \blue{}, and (c) \green{}. For each row, the left panel is the signal, the middle panel the fit result, and the right panel the error ($\mathrm{signal} - \mathrm{fit}$).}
\end{figure*}

\begin{figure*}[ht]
    \centering
\includegraphics[]{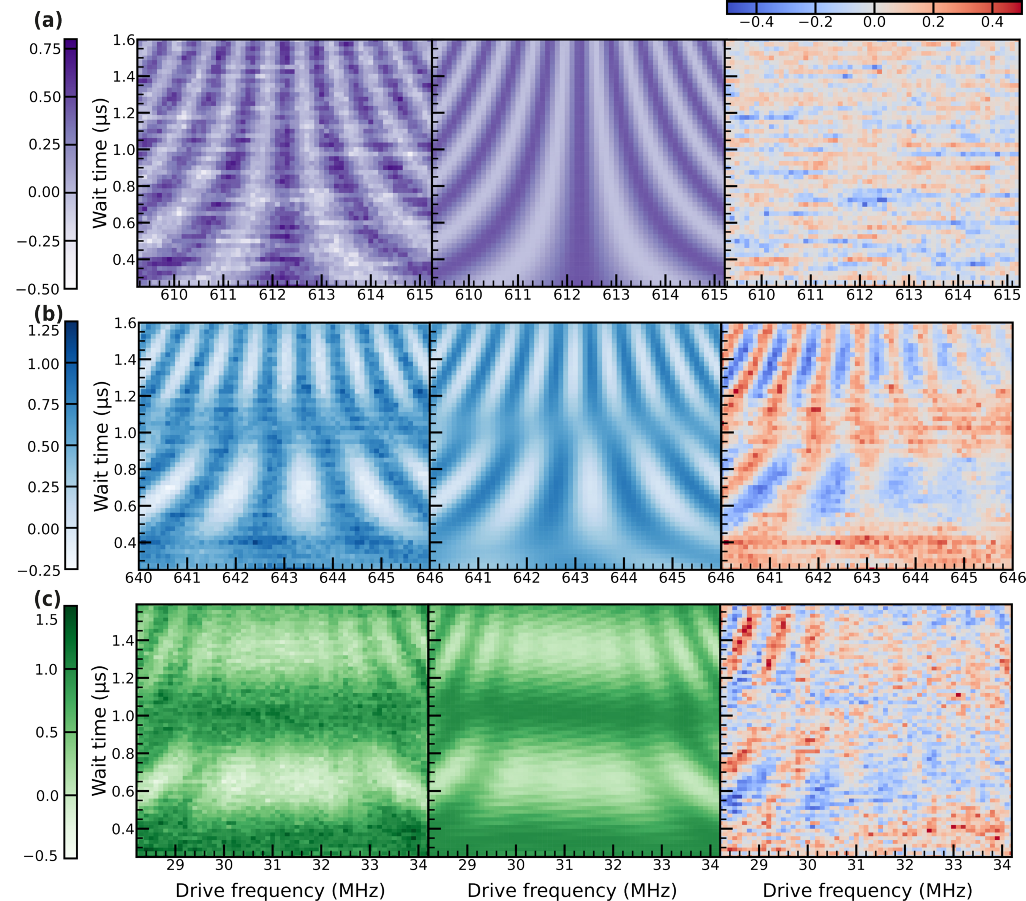}
\caption{\label{fig:ramsey_fitting}Results from the combined data fit to Rabi and Ramsey oscillations from transitions (a) \purple{}, (b) \blue{}, and (c) \green{}. For each row, the left panel is the signal, the middle panel the fit result, and the right panel the error $|\mathrm{signal} - \mathrm{fit}|$.}
\end{figure*}

To fit the Hamiltonian parameters, we fix the parameters spin-orbit coupling strength $\lambda = 830$~GHz, and ground state orbital susceptibility $q = 0.171$, based on previous measurements~\cite{Thiering2018a,Rosenthal2023,Guo2023}. 
We also assume the magnetic field $y$-component to be zero, focusing only on distinguishing between parallel and perpendicular magnetic field strengths.
Likewise, we assume the strain components $\alpha\_{Egy} = 0$, only considering the $\alpha\_{Egx}$ component. 
By simulating the full Hamiltonian in equation~\ref{eq:hamiltonian} in combination with DC and AC magnetic field terms from equation~\ref{eq:magnetic}, we are able to find fitted parameters for the ground state hyperfine coupling, strain, and magnetic field that match the Rabi and Ramsey driving data presented in Fig.~\ref{fig:spin} in the main text, as shown in Figs.~\ref{fig:rabi_fitting}-\ref{fig:ramsey_fitting}, with fitted parameters in Table~\ref{tab:parameters_app}.

Once the ground state parameters are fitted, we can estimate the excited state hyperfine parameters and strain.
From group theory, the dopant atom at the inversion center should be located at a node of the $e_u$ excited-state orbital wavefunction, and we thus expect the Fermi contact term, which depends only on the spin density at the dopant, to be zero.
We therefore assume that the hyperfine interaction in the excited state is only dependent on the strength of the dipole-dipole interaction.
This anisotropic component is limited by the \DDDD{}-symmetry of the defect to have the form $A\_{dd}=A_\parallel=-2A_\perp$~\cite{Munzarova2004}.
Using the known zero-strain hyperfine splitting of the optical transition of 453~MHz~\cite{Harris2023,Parker2023}, we estimate this parameter at -232~MHz.
We then use the 676~MHz strain splitting of the $f_1/f_2$ peaks to estimate the excited state strain at -209~MHz.
These parameters are summarized in Table~\ref{tab:parameters}.

\begin{table}
\centering
\begin{tabular}{l|c|c|c}
\makecell{\textbf{Parameter} \\ \ } & \makecell{\textbf{Symbol} \\ \ } & \makecell{\textbf{Best fit} \\ \textbf{value}} & \makecell{\textbf{Relative} \\ \textbf{fit error}} 
\\ \hline
\small
DC $B$-field & $b_{x\,\mathrm{DC}}$ 
& 
$6.03$ MHz
& 
$1 \times 10^{-4}$ 
\\
DC $B$-field & $b_{z\,\mathrm{DC}}$ 
& 
$1.55$ MHz
& 
$6 \times 10^{-4}$ 
\\
AC (drive) $B$-field & $b_{x,\mathrm{AC}}$ & 8.92 MHz & $5 \times 10^{-4}$
\\ 
AC (drive) $B$-field & $b_{z\,\mathrm{AC}}$ & 5.00 MHz & $2 \times 10^{-4}$
\\ 
Hyperfine coupling & $A_{\parallel}$ & 673.8 MHz & $4 \times 10^{-4}$
\\
Hyperfine coupling & $A_{\perp}$ & 670.95 MHz & $7 \times 10^{-5}$
\\
Strain & $\alpha_{Egx}$ & 928.4 GHz & $9 \times 10^{-4}$
\normalsize
\end{tabular}
    \caption{Fitted Hamiltonian and drive parameters.}
    \label{tab:parameters_app}

\begin{tabular}{c|c}
\textbf{Transition} & \makecell{\textbf{Estimated} \\ \textbf{frequency}}
\\ \hline 
\purple & 612.3 MHz 
\\
\blue & 642.6 MHz
\\
\green & 644.1 MHz
\end{tabular}
\caption{Transition frequencies estimated by the datafit.}
\end{table}

As discussed in the main text, the DC magnetic field we fit to match the ground state dynamics has a total magnitude of 223~$\upmu$T, which is much larger than the Earth's magnetic field at our location ($\sim55$~$\upmu$T).
The large magnetic field might be explained by magnetization of components near the sample within the cryostat.
It is also possible that other effects, such as Jahn-Teller distortion, have similar effects to a perturbing magnetic field.
Under the assumption that the residual splitting is caused by a magnetic field (or that the effective magnetic field due to other effects is the same in the ground and excited state), we estimate the difference in memory splitting between the ground and excited state at $\Delta\omega\_M=2\pi\cdot10.4$~kHz.
This is several orders of magnitude smaller than the $\Delta\omega\_M=2\pi\cdot35$~MHz value shown for \hsiv{} under large magnetic field bias~\cite{Stas2022}.
Using the optical lifetimes for \siv{} and \snv{}, equation~\ref{eq:excitation_fidelity} suggests that \hsiv{} could be excited only 2 times under the bias scheme in~\cite{Stas2022} before memory fidelity drops below 95\%, while the \hsnv{} in this paper could be excited nearly 1.5 million times (although transition cyclicity would limit fidelity well before this point).

\section{Coherence}\label{sec:coherence}
Phonon-mediated decoherence is a significant source of decoherence for the group-IV color centers~\cite{Jahnke2015}, and it is therefore important to determine whether this effect will also cause decoherence the \hsnv{} hyperfine levels.
The coherence of a transition should be dependent on a term $\lambdaeff$ equal to the difference between the transition frequency in the upper and lower ground state branch~\cite{Harris2024}.

\begin{figure*}[ht]
    \centering
    \includegraphics[width=\linewidth]{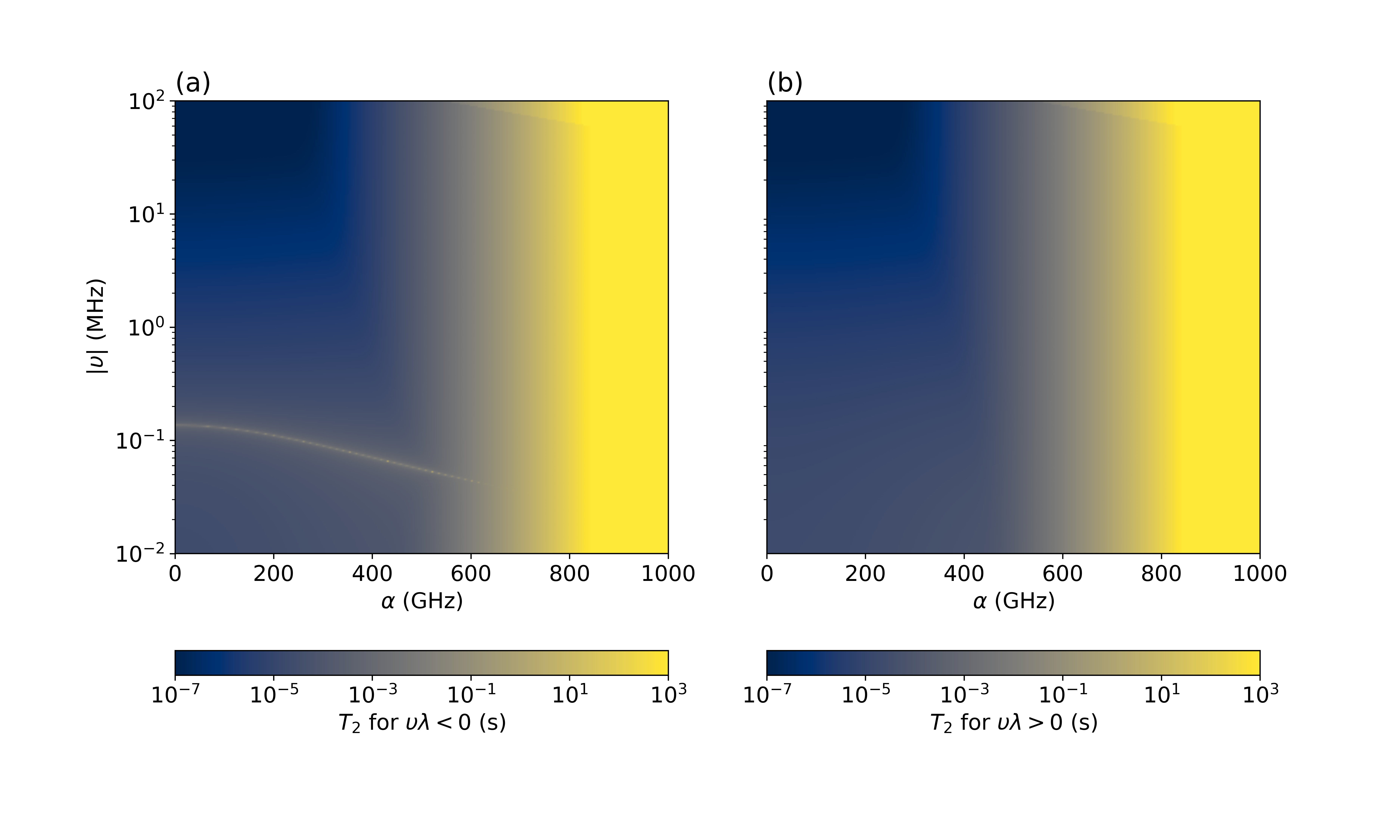}
    \caption{
    Predicted $T_2$ at 1.7~K for the broker qubit as a function of the nuclear spin orbit parameter $\upsilon$, and strain $\alpha$ assuming (a) $\upsilon\lambda < 0$ and (b) $\upsilon\lambda > 0$.
    }
    \label{fig:hyperfine_coherence}
\end{figure*}

Since there is no dependence of the broker/memory qubit energies on the branch state to first order, we expand the eigenenergies of equation~\ref{eq:hamiltonian} to second order in $A_{\perp/\parallel}/\Delta$ and $\upsilon/\Delta$.
Using the definition of broker/memory subspaces from the main text, the energy levels are then
\begin{equation}\label{eq:energies_2nd}
    \begin{split}
        E_{1\_B0\_M}^\pm=E_{1\_B1\_M}^\pm =& \frac{A_\parallel}{4} \pm \left( \frac{\Delta}{2} + \frac{\upsilon}{2}\frac{\lambda}{\Delta} \right), \\
        E_{0\_B1\_M}^\pm=& -\frac{A_\parallel}{4} + \frac{A_\perp}{2}\frac{\alpha}{\Delta} + \frac{\alpha\lambda\upsilon}{\Delta^3} \\
        & \pm \left( \frac{\Delta}{2} - \frac{\upsilon}{2}\frac{\lambda}{\Delta} + \frac{A_\perp^2}{4\Delta}\left( 1-\left(\frac{\alpha}{\Delta}\right)^2\right)  \right), \\
        E_{0\_B0\_M}^\pm=& -\frac{A_\parallel}{4} - \frac{A_\perp}{2}\frac{\alpha}{\Delta} + \frac{\alpha\lambda\upsilon}{\Delta^3} \\
        & \pm \left( \frac{\Delta}{2} - \frac{\upsilon}{2}\frac{\lambda}{\Delta} + \frac{A_\perp^2}{4\Delta}\left( 1-\left(\frac{\alpha}{\Delta}\right)^2\right)  \right)
    \end{split}
\end{equation}
where the $\pm$ corresponds to the upper/lower branch.
The transition energy of each qubit corresponds to the differences between the $\ket{0/1}$ states, and will be conditioned on the state of the other qubit.
We therefore calculate the conditional energies
\begin{equation}
    \begin{split}
        E_{\mathrm{B}\left|1\_M\right.}^\pm =& E_{1\_B1\_M}^\pm - E_{0\_B1\_M}^\pm \\
        =&\frac{A_\parallel}{2} + \frac{A_\perp}{2}\frac{\alpha}{\Delta} + \frac{\alpha\lambda\upsilon}{\Delta^3} \\
        & \pm \left( \upsilon\frac{\lambda}{\Delta} + \frac{A_\perp^2}{4\Delta}\left( 1-\left(\frac{\alpha}{\Delta}\right)^2\right) \right),\\
        E_{\mathrm{B}\left|0\_M\right.}^\pm =& E_{1\_B0\_M}^\pm - E_{0\_B0\_M}^\pm \\
        =& \frac{A_\parallel}{2} - \frac{A_\perp}{2}\frac{\alpha}{\Delta} + \frac{\alpha\lambda\upsilon}{\Delta^3} \\
        & \pm \left( \upsilon\frac{\lambda}{\Delta} + \frac{A_\perp^2}{4\Delta}\left( 1-\left(\frac{\alpha}{\Delta}\right)^2\right) \right), \\
        E_{\mathrm{M}\left|1\_B\right.}^\pm =& 0,\\
        E_{\mathrm{M}\left|1\_B\right.}^\pm =& A_\perp \frac{\alpha}{\Delta} \\
    \end{split}
\end{equation}
Taking the difference between the positive and negative branches, we can see that that $\lambda\_M=0$, and
\begin{equation}\label{eq:lambdab}
    \lambda\_B = 2\upsilon\frac{\lambda}{\Delta} + \frac{A_\perp^2}{2\Delta}\left( 1-\left(\frac{\alpha}{\Delta}\right)^2\right).
\end{equation}
We can therefore conclude that the memory qubit should be immune to the typical phonon-mediated decoherence, while the broker qubit would have a coherence time of~\cite{Harris2024}
\begin{equation}
    T_2 = \frac{4\pi}{\lambda\_B\left( 1 - \exp\left( -\frac{2\pi}{\lambda\_B\gamma} \right) \right)}
\end{equation}
where $\gamma$ is phonon scattering rate between the branches at $\pm\Delta/2$.
We show the predicted coherence as a function of the applied strain and nuclear spin-orbit coupling in Fig.~\ref{fig:hyperfine_coherence}.

Although the nuclear spin-orbit parameter $\upsilon$ has not been measured, given the $r^{-3}$ dependence on the electron density we expect its bare value to be of the same order magnitude as the dipole-dipole coupling, which we here estimate to be on the order of 1~MHz (see Table~\ref{tab:parameters} and Appendix~\ref{sec:fitting}).
It is further expected that IOC should have a Ham reduction similar to what happens for SOC~\cite{Thiering2018a}, and the final value should be of the order $\upsilon\approx0.1-1$~MHz.
Under the 928~GHz strain, we expect phonon dephasing limits coherence to $>1$~s at 1.7~K, and even with very low strain, coherence times of the order of 10-1000~$\upmu$s should be possible.
Depending on the relative sign of $\upsilon$ and $\lambda$, it may also be possible to choose a bias strain such that $\lambda\_B\rightarrow0$ in equation~\ref{eq:lambdab}, as can be seen by the narrow region of high coherence in Fig.~\ref{fig:hyperfine_coherence}a.

\end{document}